\documentclass[12pt]{article}
\usepackage{amssymb}
\usepackage{graphicx}

\newcommand{\ba}{\begin{array}}
\newcommand{\ea}{\end{array}}
\newcommand{\bd}{\begin{displaymath}}
\newcommand{\ed}{\end{displaymath}}
\newcommand{\be}{\begin{equation}}
\newcommand{\ee}{\end{equation}}
\newcommand{\bea}{\begin{eqnarray}}
\newcommand{\eea}{\end{eqnarray}}

\def\m{\mu}

\begin{document}
\title{{\begin{flushright}
\end{flushright}}\vskip 0.5 cm {\textbf {  SQCD Inflation \&  SUSY
 Breaking}}}
\author{Philippe Brax, Carlos A. Savoy and Arunansu Sil\\ \\
\small{\em Institut de Physique Theorique, CEA/Saclay, F-91191 Gif-sur-Yvette Cedex, FRANCE.}}
\date{}
\maketitle
\vskip 2cm
\begin{abstract}

A model of generalized hybrid inflation in a supersymmetric QCD theory is
proposed whose parameters are the gauge coupling and quark masses. Its 
gravitational coupling to another SQCD sector induces a metastable supersymmetry breaking
vacuum  of the ISS type as ISS quarks become massive at the end of inflation. 
Using  a known mechanism with a gravitational breaking of the baryon number and 
the gauging of flavour symmetries, we find that gauge mediation of supersymmetry breaking 
is compatible with the dynamics of the inflation sector. Reheating proceeds via the 
thermalization of the ISS messengers into the standard model states. This
setup contains a single  dimensionful parameter in the form of  a quark mass term in the
inflationary sector, i.e. all other scales involved are either related to this single mass
parameter or dynamically generated.

\end{abstract}
\maketitle
\vskip 0.7cm
\newpage

\section{Introduction}

New large effective scales are suggested  by many phenomenological approaches to physics beyond the 
Standard Model: neutrino masses, baryogenesis, FCNC, CP violations and, in particular, inflation. 
In supersymmetric theories - where the interplay between large and small scales is more comfortable 
and $M_P = 2.4 \times 10^{18}$ GeV is a natural cutoff -  the gravitational interactions in the effective supergravity lagrangians 
have been  often taken into account to generate renormalizable operators when some fields take relatively 
large vacuum expectation values (v.e.v.'s). Of particular interest are the possibilities of gravity mediation 
of symmetry breaking effects from an otherwise hidden sector to the observable world. The classical example 
is supergravity mediation of supersymmetry breaking\footnote{For a review, see, {\it e.g.} \cite{nilles}.}  
to the supersymmetric extensions of the Standard Model (SM), generically referred to here as MSSM, where 
the effective supersymmetry breaking is suppressed by a $M_P$ factor. Another kind of examples arises in 
models where the decay of the inflaton into SM states, the reheating, arises from gravitational 
operators.

In the present paper we investigate whether gravitational interactions between two hidden sectors responsible, respectively, 
for supersymmetry breaking and inflation could be relevant and whether the relatively large scales generated in supersymmetric 
inflation could be responsible for supersymmetry breaking in the other sector. This is done here in the context of two attractive 
hidden sector candidates: supersymmetry breaking in a metastable vacuum of a supersymmetric QCD (SQCD) theory\cite{ISS} and 
inflation in another SQCD model, where the peculiar properties of the inflationary model are fixed by  the choice of the 
numbers of colours and flavours. This novel inflationary model is interesting {\it per se} because its effective superpotential 
arises from a SQCD as its UV completion. We also discuss the feedback to the inflation scenario: how the reheating could proceed 
through the states of the ISS sector.

In dynamical supersymmetry breaking (DSSB) by means of the Intriligator-Seiberg-Shih mechanism\cite{ISS} (ISS), the Universe lives in the metastable vacuum of a asymptotically free SQCD  where the overall chiral symmetry is broken by a quark mass term. The resulting superpotential in the dual magnetic theory is of the O'Rafertaigh  type, where the  linear term, proportional to the quark mass, defines the supersymmetric breaking scale. The $R$-symmetry of the superpotential is only broken by its non-perturbative term that induces a supersymmetric vacuum. The (meta-)stability is ensured if the DSSB scale - hence the quark mass in the electric theory - is small enough as compared to the SQCD scale.

The effective supersymmetry breaking in the standard supersymmetric gauge and matter sector (MSSM) requires a mediation mechanism of the DSSB from the otherwise hidden ISS sector.  However, in its original version, ISS does not fulfill the well-known requirements to implement gaugino masses,namely, $R$-symmetry breaking\cite{nelsei} and scalars with charges $R \ne 0, 2\,$\cite{shih}. The original ISS picture must be enriched\cite{Rbkg} and we follow here the elegant suggestion of \cite{abel1} and introduce a baryonic term in the superpotential to fix the magnetic quarks to have charges  $R=1$ and to produce spontaneous $R$-symmetry breaking. Then, it is possible to identify a subgroup of the flavour symmetry with the GUT $SU(5)$ group (or the corresponding SM gauge symmetry subgroups). The magnetic quarks become plausible messenger candidates for gauge mediation of DSSB (GMSB)\footnote{For a review, see, {\it {e.g.}} \cite{GMSB}.} as confirmed by a phenomenological analysis\cite{abel2}, which finds, in particular, that $O(\mathrm{TeV} )$ gaugino masses, require a relatively low supersymmetry breaking scale, $O(10^8\, \mathrm{GeV})$.

Although quite attractive, the resulting model  has two free scale parameters much smaller than the scale associated to the gauge coupling: the DSSB one - related to the electric quark mass in ISS - and the coefficient of the baryonic term, which controls the $R$-symmetry breaking. Furthermore, in the effective magnetic theory both parameters must be of the same order of magnitude to produce realistic mass spectra in the MSSM through gauge mediation\cite{abel2}. These low scales are protected by global symmetries, equivalent to a quark chirality and a baryon-number and we would like here to propose a possible origin for their breaking at relatively low scales.

In recent works\cite{ss, shift}, the authors  have shown that the scales of inflation and supersymmetry breaking in a metastable vacuum can be naturally connected from the assumption that the two sectors are coupled only by (super)gravity.  Of course, for the suggestion to be meaningful, the inflationary model must be also endowed with some basic properties of the ISS model:  supersymmetry, $R$-symmetry and, possibly, other global symmetries, which should also be present in the gravitational interactions. As  a  case study for providing such relations between inflation and supersymmetry breaking,  supersymmetric hybrid inflation was discussed in \cite{ss}: the scale of hybrid inflation is constrained by experimental data to be very large and the resulting supersymmetry breaking is also large, just consistent with supergravity mediation. Then,  with a new model for inflation, which includes a shift symmetric K\"ahler potential along with an inflationary potential of the hilltop type, and leads for a much lower scale for inflation, we showed in \cite{shift}  that the scale of supersymmetry breaking comes out proportionally smaller\footnote{A different inflation scenario where the inflaton 
rolls down to the ISS metastable vacuum has been proposed in \cite{craig}.}.

In this paper, we want to investigate a similar scenario with a further motivation to include the $R$-symmetry breaking which is necessary for making gauginos massive. Since in this scheme the DSSB scale is related to the inflation scale, we need an inflation model such that the ISS effective scales come out naturally low. To implement this requirement and to carry on the parallelism with the ISS model and, also the MSSM, we build a SQCD inflationary model where, apart from the dynamically generated scale,  the dimensional parameters have a UV interpretation in terms of quark masses. Besides this theoretical motivation, the model turns out to be quite consistent with inflation data. We show that, at the end of inflation, its $R$-symmetric gravitational couplings with the fields in the ISS sector can generate  low scale DSSB and $R$-symmetry breaking, suitable for gauge mediation.

For not too small quark masses in the inflation SQCD, the mesons in its IR phase are massive enough for the inflaton to mainly decay into ISS magnetic quarks, in particular those coupled  to the ordinary matter through the SM gauge sector which play the role of messengers in the model. The calculated reheating temperature is quite acceptable. Therefore, in this scheme, one has a sort of ``ISS mediation'' of the inflaton energy to the SM or,  more precisely, the MSSM particles.

As a step further, we suggest an upgrade of our inflationary model where the dimensional parameters in the latter are obtained as quark masses of the former. This is done by increasing the number of flavours by one unit, from $N_f = N$ to  $N_f = N + 1$, with one heavier quark whose mass then corresponds to the second parameter needed in the smooth hybrid model, so completing our SQCD picture of inflation.

The paper is organized as follows. The basic assumptions are presented in section II and are followed, in section III, by a brief review of the deformed ISS model of \cite{abel1} including the phenomenological constraints on both parameters in the effective O'Rafertaigh superpotential. The SQCD inflationary model is described  and confronted to data in section IV to obtain the constraints on the parameters. It is then coupled through supergravity to the ISS fields in section V to produce the deformation of the ISS model with the low scales in the magnetic phase fixed up to $O(1)$ factors  by both the dynamical SQCD scales involved.  Section VI explains how the ordinary matter and gauge particles are produced through the production and decays of the GMSB messengers in the ISS sector during the reheating period. The last section summarizes our conclusions.

\section{The basic setup: a 3-fold Universe}

As in \cite{ss}, the set-up consists of three components namely, the inflationary sector (Infl), the supersymmetry breaking sector (here the deformed ISS sector, dISS) and, of course, the MSSM sector.

Fields in different sectors have only (super)gravitational interactions described by an effective supergravity theory where the superpotential and a K\"ahler potential respect the symmetries of the different sectors. Each sector is phenomenologically very constrained. For the MSSM we only consider here those related to scalar and gaugino masses. We want to study the possible interferences between the phenomenologies in the different sector due to those gravitational interactions. They are generically present in supergravity and are restricted by some symmetries common to both sectors, {\it e.g.}, $R$-symmetry. Actually,  the connection between the potential flatness and $R$-symmetry makes its choice rather natural as flatness is the basic idea in hybrid inflation models as well as in building ISS metastable vacua. They could break less essential accidental symmetries, {\it e.g.,} chiral symmetries broken by quark mass terms.

 However we are forced to break these seclusion rules in one respect. Indeed  DSSB must be mediated from dISS to MSSM. Gravity mediation was envisaged in \cite{ss}, but the fundamental scales come out all very high, the gaugino masses were not accounted for in that first analysis. In GMSB one assumes that gauge multiplets are coupled to both sectors  and the minimal choice\cite{abel1} is that the SM gauge particles couple to some of the (so-called) quarks in the magnetic phase of the dISS sector which then become messengers. Therefore the SM gauge symmetries must be embedded into the ISS chiral flavour group (in the present paper in a vector-like way). This model for GMSB with ISS is also adopted here. The consequences have been widely discussed in the literature\cite{abel2} and are summarized below. In our setup, it could lead to a novel reheating mechanism.

As mentioned in the Introduction, we assume here that each sector is described by its own SQCD theory, with its gauge sector, its chiral multiplets  called quarks and, by dimensional transmutation, a characteristic scale. In practice, we take simpler (in particular simple) gauge groups and representations than in the MSSM, basically one $SU(N)$ with $N_f$ vector-like pairs of  fundamental representations for the chiral multiplets, or ``quarks''. For ISS, the IR or magnetic theory with a metastable vacuum has a known  UV completion\footnote{For a review, see, {\it e.g., } \cite{seiberg95}.}, an (electric) asymptotically free SQCD if it satisfies $N+1 \le N_f \le \frac{3}{2}N$. The implementation of $R$-symmetry breaking as in \cite{abel1} then requires $N=5$ and  $N_f=7$ where the $SU(7)$ factor in the flavour chiral symmetry is explicitly broken by a baryonic term to $SU(5) \otimes SU(2)$.

SQCD with $N=N_f=2$ has been discussed in the literature\cite{dvali} because its low energy superpotential is of the supersymmetric hybrid inflation type\cite{hybrid} with a flat inflaton direction in the potential. It leads to a less satisfactory model of inflation and a large supersymmetry breaking\cite{ss} from its gravitational coupling to ISS.  Here we alter the inflation scenario by increasing the number of colours, $N$ and also by considering $N_f=N+1$ SQCD inflation. For $N=4$, the low energy effective superpotential in terms of so-called mesons and baryons has a similarity with the so-called smooth hybrid inflation model (SHI)\cite{smooth}.

However,  there are two key differences: {\it a)} there are two waterfall fields in SHI to be compared to the $N^2-1$ mesons in  $N=N_f$ SQCD, which in the vanishing quark mass limit are massless goldstone bosons; {\it b)} SHI has two free scales, a cutoff $\Lambda_0$ somewhat below $M_P$ and the inflation scale $\Lambda_{\mathrm{eff}}^2$ which in our model are defined as the strong coupling scale and $m_Q \Lambda_0$, respectively, where $m_Q $ is the mass of the additional quark in the UV region. The addition of this quark promotes the theory to the $N_f=N+1$ theories. The SHI  superpotential and its free parameters are justified as the IR dynamics of a SQCD theory. Actually, this fact is the main motivation (on the inflation side) for the new inflationary model proposed in section VII.

The $R$-symmetric supergravity coupling between the dISS and the Infl mesons which produce a linear term in the ISS superpotential, namely, the mass term for the electric quarks that induces DSSB. The DSSB scale is then controlled by the two inflation parameters. The CMB data leaves only one free parameter: the inflation scale can be tuned down by lowering the scale $\Lambda_0$ without affecting the agreement with data. Because it also controls the DSSB scale in the dISS sector, this one can be lowered as well. As a result, $\Lambda_0$ turns out to be fixed by the feedback from the MSSM phenomenology after GMSB. This shows a tight connexion between the dISS and Infl phenomenologies in our scenario.

 We also find a new reheating mechanism. Reheating presupposes some link between the Infl and the MSSM sectors. This was possible in \cite{ss} by a gravitational coupling of the waterfall fields to right-handed neutrinos. In the present case this coupling is reduced by the required symmetries. Nevertheless, the inflaton field now decays into dISS quarks and the MSSM particles are produced by thermalization since the two sectors intersect  through the MSSM gauge fields. Reheating is mediated by ISS quarks.


\section{ISS sector: supersymmetry breaking in a SQCD metastable vacuum}

In this section we briefly review the ISS approach and the modification suggested in \cite{abel1}. In the UV region, the dISS sector corresponds to an asymptotic free SQCD  with $N = 5$ colours and $N_f = 7$ flavours for the quark superfields,  $Q_i^{\alpha}\, , \tilde{Q}_i^{\alpha}\,$ (the electric theory) where $i=1,\dots,7$ and $\alpha =1,\dots,5$ . The GMSB kit, consists of flavoured particles in  the dISS sector as messengers that couple to the MSSM  only through these gauge interactions: a subgroup of the flavour group containing the SM group is to be gauged. For simplicity we gauge a  whole $SU(5)$ and identify it with the GUT $SU(5)$.  The chiral $SU(7)$ flavour symmetries  are then broken to $SU(2)\otimes SU(5)$  by construction. For simplicity we still refer to the gauged $SU(5)$ as a flavour group in this section.

Below the strong coupling scale $ \Lambda_s$  the model has a dual description in terms of a IR free magnetic theory with magnetic gauge group $SU(2)$ and $N_f = 7$ quarks and antiquarks, $q_i^a\, , \tilde{q}_i^a\,$, where $a=1,2$ denotes the magnetic colours,  together with the so-called mesons which define  a matrix $\Phi_{ij}$. The matching of the degrees of freedom and of the preserved flavour symmetries is ensured by a superpotential  in the magnetic theory. The deformed ISS model has  two more terms in the superpotential and reads:
\be
W_{dISS} = \Phi_{ij} q_i \tilde q_j -  \mu_{ij}^2 \Phi_{ji} +
m_q \epsilon_{ab} \epsilon_{rs} q^{a}_r q^{b}_s +
(N_f - N)\left(h^{N_f} \frac{{\rm {det}} \Phi}{\Lambda_s^{3N - 2N_f}}\right)^{\frac{1}{N_f - N}}
\label{WdISS}\, ,
\ee
where $\mu^2_{ij} =  {\rm diag} (\mu_2^2,\mu_2^2,\mu_5^2 \dots \mu_5^2)$, and the indices $r\,,s=1\,,2$ denote the quarks that are singlets under the GUT  $SU(5)$. The first and last term constitute the superpotential of the magnetic dual of the UV SQCD. The second term was introduced in ISS and amounts to a breaking of two chiral symmetries: it is responsible for the supersymmetry breaking in a metastable vacuum at a scale proportional to  the parameters $\mu_2$ and/or $\mu_5$.

The third term that ``deforms'' the ISS model, behaves as a $SU(2)$ baryon, hence it breaks a baryon number and leads to the dynamical breaking of $R-$symmetry. Indeed, if one neglects the last term, for $\Phi \ll \Lambda_s$, the remaining terms have all  $R = 2$ if one defines the R-charges of $\Phi\,, q$ and $\tilde q$ as $2, 1$ and $-1$, respectively (in the original ISS model, the latter can be taken to be 0). Therefore the  $R-$symmetry can be broken  at a scale $O(m_q)$. Actually, this baryon term introduces a runaway direction towards a non-supersymmetric vacuum at infinity, but the potential is stabilized by the Coleman-Weinberg radiative corrections, proportional to the supersymmetry breaking parameters, $\mu_2$ and/or $\mu_5$.

It was shown\cite{abel1}  that the introduction of this term in Eq.(\ref{WdISS}) of the magnetic dual theory is responsible for shifting $\langle \Phi \rangle$ away  from zero,  thereby breaking the $U(1)_R$ . Without going into many details \cite{abel1}, we need to recall a few  points that are crucial in our discussion. It has been shown that the potential is minimized when  $\Phi$ is diagonal and does not break $SU(2)\otimes SU(5)$, its v.e.v.'s being $O(m_q)$, while only the $SU(5)$ singlet quarks get a v.e.v so that the SM (here the GUT) symmetry is preserved as it must be.  With some technically natural choice of the parameters,
\begin{equation}
\mu_2 \simeq \mu_5 = O(m_q)\, ,     \label{condition}
\end{equation}
one obtains a viable model with DSSB and enough $R-$symmetry breaking by fields with $R \ne 0,2$ and coupled to the $SU(5)$ as required to give masses to gauginos and implement GMSB. The mass degeneracy of the dISS squarks with non-trivial SM quantum  numbers is broken giving masses to the MSSM  gauginos. The requirement that gaugino masses are   of order $O(\mathrm{GeV})$ leads to  the constraint $\mu_5 = O(10^8\, \mathrm{GeV})$.

Our goal is to provide a mechanism to generate these parameters and reproduce those constraints from the coupling the model for inflation. For an analysis of the dISS phenomenology see \cite{abel2}.

\section{Inflation sector: a SQCD model}

We discuss here an inflationary model  represented by a strongly coupled supersymmetric $SU({\cal N})$ gauge group with ${\cal N}_f ={\cal N}$ flavours of quark superfields $\mathcal{Q}_{i}$ and $\bar \mathcal{Q}_{i}$ $(i = 1, \dots {\cal N}_f)$ in the ${\cal N}$ and $\bar {\cal N}$ representations of the gauge group. The system has a non-anomalous global symmetry $G = SU({\cal N}) \otimes SU({\cal N}) \otimes U_B(1) \otimes U_R(1)$. It turns out that the case  ${\cal N}=4$, which is a SQCD generalization of  the smooth hybrid inflation model\cite{smooth, smoothshafi}, fits better the data and we concentrate on it from now on.

Below the scale $\Lambda_0$ (where the $SU({4})$ gauge coupling becomes large)\footnote{More precisely (see below) $\Lambda_0$ is the strong coupling scale of the 
parent SQCD theory with one more flavour.},
the theory  is  described by an effective theory of composite mesons in the representation $(\underline{4},\bar{\underline{4}})$ of the chiral flavour group, one baryon and one anti-baryon,
\begin{equation}
T_{i j}=  \frac{1}{2}\Lambda_0^{-1} \mathcal{Q}^{a}_{i}  \bar{\mathcal{ Q}}^{a}_{i}\,,  \qquad
B= \frac{1}{3}\Lambda_0^{-3}\epsilon_{ijkl} \mathcal{Q}^{1}_{i}  \mathcal{Q}^{2}_{j}  \mathcal{Q}^{3}_{k}  \mathcal{Q}^{4}_{\ell} ,  \qquad
\bar B = \frac{1}{3}\Lambda_0^{-3} \epsilon_{1234}{\bar \mathcal{ Q}}^{1}_{i}
{\bar \mathcal{Q}}^{2}_{j}{\bar \mathcal{ Q}}^{3}_{k}{\bar \mathcal{ Q}}^{4}_{l}, \label{hadrons}
\end{equation}
where the superscripts are colour indices. All these fields have $R=0$ and another field, denoted by $S$, must be introduced to implement the charge $R=2$ for the superpotential, which is then fixed as the flavour symmetry invariant\footnote{ For reviews see, {\it e.g.},\cite{seiberg95}. The relationship between the existence of the superpotential, and the matching of anomalies and degrees of freedom is also discussed in \cite{BGS} on more general grounds.}
\be
W_{Infl} = S \left ( \frac{{\rm det}T}{\Lambda^2_0} - {B \bar B} - \Lambda^2_{\rm eff} \right ),
\label{wx}
\ee
Notice the presence of an additional scale, $ \Lambda_{\rm eff}$  that we treat here as a parameter and refer to \cite{seiberg95} and section VI for a discussion of its meaning.

In the so-called meson branch of the theory, where T does not vanish, it can be represented in terms of a non-linear realization by  Nambu-Goldstone bosons (NGB)  of the global symmetry $SU(4) \times SU(4)$ broken down to $SU(4)_V$ as follows,
\be
T = \chi \exp{\left( i \frac{t^{\alpha}  \lambda^{\alpha}} {\langle \chi \rangle} \right)}, \label{NGB}
\ee
with $\alpha = 1, ..., 15$ where $t^{\alpha} $ represents the NGB superfields and $\lambda^{\alpha}$ the $SU(4)$ generators. Obviously these NGB are cosmologically relevant and will be treated in the reheating section. The superpotential for $B=\bar{B} = 0$ and replacing ${\rm {det}}T = \chi^4$ in Eq.(\ref{wx})  coincides with the smooth hybrid inflation one, namely,
\begin{equation}
W_{Infl} = S \left ( \frac{\chi^4}{\Lambda^2_0}  - \Lambda^2_{\rm eff} \right ),
\label{infsup}
\end{equation}
which has a supersymmetric minimum at
\be
 \langle \chi \rangle = (\Lambda_{\rm eff} \Lambda_0)^{1/2},  \qquad \qquad S = 0 \, .
\label{vev1}
\ee

But there is a relevant difference between the two model besides the presence of the NGB: the scales in Eq.(\ref{infsup}) have now a physical meaning as the gauge theory  scale for the cutoff $\Lambda_0$ and the scale $\Lambda_{\rm eff}$ can be related to the explicit breaking of a chiral symmetry if one massive flavour is added to the  ${\cal N}_f ={\cal N}$ SQCD\cite{seiberg95}.  To see it, let us start with a ${\cal N}_f ={\cal N}+1=5$ SQCD, and add a superpotential corresponding to a quark mass term,
\begin{equation}
W_m = \mathrm{Tr } \hat{m} \mathcal{Q} \bar{\mathcal{ Q}} \label{wm},
\end{equation}
where the trace is taken over the five flavours and four colours. At low energies this theory is better formulated in terms of the mesons and baryons which are defined analogously to Eq.(\ref{hadrons}) but, since now $i=1 ,\,\dots , \,5\,,$  the baryons carry a free flavour index, $\hat{B}^i$ and the meson matrix $\hat{T}_{ij}$ is correspondingly larger. The low energy superpotential is,
\be
\hat{W} = {\hat B} {\hat T} {\hat {\bar B}} -  \frac{{\rm det} {\hat T}}{\Lambda^{2}_0}
+ {\Lambda}_0{\rm {Tr}} \hat{m}\ {\hat T},
\label{w1}
\ee
where $\Lambda_0$ is the strong coupling scale of the ${\cal N}_f$ = $\cal N$ +1 SQCD and the last term  is the counterpart of  Eq.(\ref{wm}).

The $R$-symmetry, that plays an important r\^{o}le for the flatness of the potential, is preserved if one chooses $\hat{m}_Q = {\rm {diag}}(0, 0,0 0, m_Q)$ corresponding 
to only one massive quark\footnote{Though in section VI, we shall complete the model by adding additional masses to the remaining quarks, hence to the associated goldstone bosons.}. The heavy degrees of freedom can be integrated out to define an effective theory at low energies (the descent relation discussed, {\it e.g.}, in \cite{seiberg95}) in terms of the baryons $B = \hat{B}^5\, , \bar{B} = \bar{\hat{B}^5}\,$,  $T_{ij} = \hat{T}_{ij}$ for $i=1,2,3,4,$ and $S=\hat{T}_{55}$. The effective superpotential coincides with Eq.(\ref{wx}) with the identification
\begin{equation}
\Lambda^2_{\rm eff} = m_Q\Lambda_0 \label{descent}\, ,
\end{equation}

Therefore the smooth hybrid inflation superpotential can be dynamically generated from a UV gauge theory completion rather than justified by {\it ad hoc} global symmetries. The cutoff scale is defined from the gauge coupling, while  the vacuum energy during inflation, $\Lambda^4_{\rm eff}$ is determined by a quark mass. Furthermore, the power of the waterfall field $\chi$ in Eq.(\ref{infsup}) is given by the number of colours, $\mathcal{N}$.

Now we turn to discuss how the scales $\Lambda_0$ and $\Lambda_\mathrm{eff}$  are  constrained
by the Cosmic Microwave Background (CMB) experimental data\cite{cobe} and WMAP\cite{wmap5}. 
We start with the superpotential in Eq.(\ref{infsup}) with $B = \bar B = 0$, the 
so-called meson branch, for simplicity.
The scalar potential in terms of the real normalized fields $\sigma = \sqrt{2} \Re(S)$ and
$\xi = \sqrt{2} \Re(\chi)$ is given by
\be
V_{Inf} (\sigma, \phi) = \left (\frac{\xi^4}{4 \Lambda_0^4} - \Lambda^2_{\rm eff} \right )^2 + \frac{\sigma^2
\xi^6}{\Lambda_0^4},
\ee
where scalar components are described by the same notations as superfields.
Following the standard smooth hybrid inflation\cite{smooth}, we see here
that the flat direction at $\xi = 0$
is now a local maximum for all values of $\sigma$
and there are  two symmetric valleys of minima present at
$\xi \simeq \pm \frac{\Lambda_{\rm eff} \Lambda_0}{\sqrt{3}\sigma}$.

An interesting point is  that the valleys contain the global SUSY minimum which lie at
$\xi = \sqrt{2\Lambda_{\rm eff} \Lambda_0}$, $\sigma = 0$ and have a slope which can actually drive the
inflaton $\sigma$ towards the right vacuum.
The potential along this valley is
\be
V (\sigma) \simeq \Lambda^4_{\rm eff} \left (1 - \frac{1}{54} \frac{\Lambda^2_{\rm eff} \Lambda_0^2}{\sigma^4} + \dots \right ),
 ~~{\rm {for}} ~ \sigma \gg \sqrt{\Lambda_{\rm eff} \Lambda_0}.
\ee
We identify  the slow roll parameters
\bea
\epsilon = & \frac{M^2_{P}}{2} \left ( \frac{V'(\sigma)}{V(\sigma)}\right )^2 \simeq & \frac{2}{729} \left (
\frac{\Lambda_{\rm eff}^2 \Lambda_0^2 M_{P}}{\sigma^5} \right )^2,\\
|\eta| & = M^2_{P}|\frac{V''(\sigma)}{V(\sigma)}| \simeq &
\frac{10}{27 \pi} \frac{ \Lambda_{\rm eff}^2 \Lambda^2_0 M^2_{P}}{\sigma^6},
\eea
where we have used the approximated
$V$  with $\sigma \gg \sqrt{\Lambda_{\rm eff} \Lambda_0}$.
The number of $e$-folds during inflation
and the temperature fluctuation are estimated to be
\bea
\label{nl}
 N_l = & \frac{1}{M^2_{P}} \int ^{\sigma_l}_{\sigma_0}
\frac{V(\sigma) d\sigma}{V'(\sigma)} \simeq & \frac{5}{6|\eta|}, \\
\label{dtt}
\Delta ~ = ~\left ( \frac{\delta T}{T} \right )_Q = & \left ( \frac{32 \pi}{45} \right )^{1/2} \frac{V^{3/2}(\sigma)}{V'(\sigma) M^3_{P}}
\simeq & \left ( \frac{9}{8\sqrt{5}\pi} \right ) \frac{1}{M^3_{P}} \frac{\sigma_l^5}{\Lambda_0^2},
\eea
where $\sigma_l$ and $\sigma_0$ indicate the values of the inflaton field when the `comoving' scale $l$ crossed outside the event horizon and the end of inflation (corresponding to the slow roll parameter, $|\eta| =1$) respectively. 
Using these parameters, the spectral index of density fluctuations, $n_s$ is estimated as
\be
\label{ns}
n_s = 1 - 6\epsilon + 2\eta \simeq 1 - \frac{5}{3 N_l} \simeq 0.97,
\ee
for $N_l \sim 56$.

We use the data from COBE\cite{cobe}($\delta T/T \simeq 6.6 \times 10^{-6}$) and the WMAP 5  results\cite{wmap5}
($n_s \simeq 0.964 \pm 0.014$) to fix the scales.
The value of $n_s$ of Eq.(\ref{ns}) is in agreement with the experimental result. 
From Eqs.(\ref{nl}, \ref{dtt}), we can write the scale $\Lambda_{\rm eff}$, or equivalently, $m_Q$, in terms of $\Lambda_0$ as follows,
\be
\Lambda_{\rm eff} \simeq 0.6 \Delta^{3/5} {\Lambda_0}^{1/5} {M_P}^{4/5}, \qquad \qquad
m_Q   \simeq  2.2 \times 10^{-7}\, \left(  \frac{M_P}{\Lambda_0} \right)^{3/5} M_P, \label{inflrelatn}
\ee
where we put $N_l = 57$. One important point should be observed. The $\epsilon$ parameter at the time of horizon exit is given by
\be
\epsilon (\sigma_l) = \frac{1}{72}\frac{1}{N^2_l} \Bigl (\frac{\sigma_l}{M_P}\Bigr)^2 \lesssim 10^{-5},
\ee
it depends upon $\sigma_l$ which is not fixed by the present data. We will use this freedom, in the next section, 
to relate the inflationary scale with the scale of supersymmetry breaking once we have fixed  the interaction between the two sectors.

So far we have neglected the supergravity corrections. With a canonical Kahler potential, the effective scalar potential 
for $\sigma$ in supergravity is given by
\be
V(\sigma) = \Lambda^4_{\rm eff}\left [ 1 - \frac{1}{54} \frac{\Lambda^2_{\rm eff} \Lambda^2_0}{\sigma^4} + \frac{\sigma^4}{M^4_P}
\right ] ~~{\rm {for}} ~\sigma \gg \sqrt{\Lambda_{\rm eff} \Lambda_0}.
\ee
As long as $\sigma \ll M_P$, the inflationary dynamics are dominated by the false vacuum energy density
$\sim \Lambda^4_{\rm eff}$ and the supergravity corrections do not modify it much\cite{smoothshafi}.
However the supergravity correction during inflation is important for the fields in the dISS sector as we discuss now.
The superfields $\Phi, q, \tilde q$ all get Hubble induced mass-square terms $\sim H^2$ and thereby settles to zero.
In \cite{ss}, this coincides with the supersymmetry breaking minimum for $\Phi$.
Here during inflation $\mu$ is very small but nonzero. As soon as $H$ is decreasing after inflation and
the Coleman-Weinberg correction stabilizes the runaway direction towards supersymmetry breaking minimum of the
dISS sector, the component fields of $\Phi$ will roll down to their minima since those are close to the
origin rather than to roll towards the supersymmetric minimum which is far away.
At the end of inflation, the inflaton field performs damped oscillations about the supersymmetric minimum
of the inflationary sector and decays. We will discuss this part in section VI.

\section{Supersymmetry breaking as a remnant of inflation}

We now turn to discuss how the parameters $\mu^2$, responsible for spontaneous supersymmetry breaking, and $m_q$, which drives $R$-symmetry breaking, could originate from gravitational interactions in the UV superpotential. In particular, how terms suppressed by inverse powers of $M_P$ can couple a pure SQCD in the  dISS sector to the 
SQCD fields of the Infl sector to produce the superpotential of the ISS model.

These couplings are controlled by the $R$-symmetry present in both sectors. In the Infl sector, the $R$-charges are only fixed as $2,\, 0,1-x,\, 1+x\, (\forall x)$, for $S,\, T, \mathcal{Q} $ and $\bar{\mathcal{Q}} $, respectively. In the original ISS model, the fields $\Phi, q, \bar{q}$ have $R$-charges $2,\, r,\, -r,$ where $r=0$ can be chosen, so that $R$ is unbroken at the metastable minimum, while the R-charges of the UV fields $Q,\, \bar{Q}$ are not uniquely fixed. In the dISS generalization, the presence of the baryon term in Eq.(\ref{WdISS}) fixes $r=1$. Then, identifying the IR baryons (two quarks $q$) to the UV ones (five quarks $Q$), one finds $R=\frac{2}{5}$ for $Q$'s and, from $R=2$ for $\Phi$, one gets  $R=\frac{8}{5}$ for $\bar{Q}$'s.

Let us now construct new UV superpotential interactions with $R=2$. Assuming the explicit breaking of the baryon number in dISS by the third term in Eq.(\ref{WdISS}), $m_qq\epsilon q$ one can write its $R=2$ avatar in the UV completion as
\be
W_{BV} = \frac{1}{M_P^2}\epsilon_{ijk\ell m} {Q}^{1}_{i} {Q}^{2}_{j} {Q}^{3}_{k} {Q}^{4}_{\ell}{Q}^{5}_{\m},
\end{equation}
The supergravity cutoff $O(M_P)$, characterizes a gravitational coupling, the only one allowed at low orders and generically present in the superpotential unless the baryon symmetry is imposed. This gives the relation
\begin{equation}
m_q = O\left( \frac{\Lambda_s^3}{M_P^2}\right) \, . \label{mq}
\end{equation}
which fixes $\Lambda_s$ once $m_q$ is fixed by the MSSM phenomenology. Therefore the $R-$symmetry breaking is controlled by the scale associated to the dISS coupling, $\Lambda_s$.

Now we turn to the main point of this section, the generation of the DSSB scale from its gravitational coupling to the inflation sector. We take for  granted that the dISS sector is secluded enough so that quark chiral symmetries are not produced from supergravity couplings or other sources\footnote{ The quark-antiquark representation being vector-like with respect to the gauged flavour symmetries, the GUT or SM ones, this is an open issue, analogous to the $\mu$-problem, and we are making a similar assumption.}. 
Therefore, in contrast to  the original ISS model\cite{ISS}, we do not assume any explicit mass terms for the quarks in the UV completion. Instead,  as in \cite{ss}, we consider  that the two sectors, dISS and Infl,  communicate only via gravity. The lowest dimensional UV term which  respects $U(1)_R$ and the other chiral $SU(4) \otimes SU(4)$  flavour symmetries in the inflation sector is given by\footnote{One can think of Tr$T Q \tilde Q$ term instead of the determinant. But this would not be invariant under the bigger symmetry group, $SU(4) \otimes SU(4) \otimes U_B(1) \otimes U_{R}(1)$. Chiral symmetry breaking in the inflationary SQCD will be introduced in the next section, but appear at a much lower scale.}
\be
W_{int} = \frac{{\rm {det}  \mathcal{Q} \bar{\mathcal{ Q}}}}{M_{P}^7}  {\rm {Tr}} f Q \tilde Q\,\quad \longrightarrow
\quad \frac{ \Lambda_0^4 \Lambda_s}{M_{P}^7} \rm{det} T\, \mathrm{Tr} \Phi, \label{wint}
\ee
where the IR avatar of the UV operator is also indicated by an arrow. Once the field ${\rm {det}T}$ gets a vacuum expectation value from Eq.(\ref{vev1})  at the end of inflation,  $W_{int}$  generates the linear terms  in the dISS superpotential and gives for their coefficients:
\begin{equation}
\mu_i^2 = f_i \frac{\Lambda_{\rm eff}^2  \Lambda_0^6 \Lambda_s}{M_{P}^7} =
O\left( \frac{ \Lambda_0^7}{M_{P}^7} \right) \Lambda_s m_Q  =
O\left(  10^{-7}\right)\left( \frac{ \Lambda_0}{M_{P}} \right)^{27/5}  \Lambda_0 \Lambda_s \, . \label{mu}
\end{equation}
where we have inserted the phenomenological  constraint from Eq.(\ref{inflrelatn}) between the two inflation parameters to obtain the last equality.

The phenomenological aspect of the model follows immediately since in order to have the right amount of gauge mediation from the 
dISS sector to the MSSM sector we must impose  that both the scales of supersymmetry ($\mu$) and $R$-symmetry ($m_q$) breaking  should be $O(10^8\,$GeV). 
This fixes the strong coupling scales,  $\Lambda_0$ and $\Lambda_s$, of both the Infl and dISS sectors. 
Note that $\sigma_l$ is $O(10^{16-17} {\rm {GeV}})$ and therefore $\epsilon$ 
turns out to be $O(10^{-8}))$.

In table 1, we have summarized the different scales involved in the problem. We have given two examples corresponding to two values of the 
supersymmetry breaking scale. It should be noted that the scale $\Lambda_0$ is larger than $\Lambda_s$. Inflation must be valid to 
higher energies than the ISS sector as it gives a mass to the electric quarks, this mass becoming the effective supersymmetry breaking 
scale at low energy. In a sense, inflation is a precursor to supersymmetry breaking. As the inflation scale $\Lambda_0$ is close to the 
Planck scale, its UV completion is a gauge theory whose domain of validity must be compatible with physics at energies close to the Planck 
scale. One enticing possibility would be to realise the inflation sector in a brane construction\cite{kut2}. The same could also be true 
of the ISS sector\cite{kut1}. In this case, the coupling between the
inflation and the ISS sector could be understood as springing from gravitational effects in the bulk. The construction of explicit brane 
models is of course beyond the scope of the present paper.

\begin{table}[t]
\begin{center}
\begin{tabular}{|c|c|c|c|c|c|c|c|}
\hline
$\mu$ & $\Lambda_0$ & $m_Q$ & $\Lambda_s$ & $\langle \chi \rangle$
& $\Lambda_{\rm eff}$ & $T_R$\\
\hline
$2.5 \times 10^8$ & $7.1 \times 10^{16}$ & $4.3 \times 10^{12}$ & $7.6\times 10^{14}$ & $6.2 \times 10^{15}$ & $5.5 \times 10^{14}$ & 100\\
\hline
$10^{9}$ & $ 10^{17}$ & $3.5 \times 10^{12}$ & $1.2 \times 10^{15}$ & $7.75 \times 10^{15}$ & $5.9 \times 10^{14}$ & 1330\\
\hline
\end{tabular}
\end{center}
\caption{\small Different scales involved in the scenario in units of GeV.}
\end{table}

\section{Reheating from gauge mediation}
Let us now discuss the inflaton decay and reheating. The inflaton fields smoothly enter an era of damped
oscillation about the supersymmetric vacuum of the Infl sector. The oscillating system
has a common mass $m_{\rm {inf}} = 2 \sqrt{2} \frac{\Lambda_{\rm eff}^2}{\langle \chi \rangle}$
and will decay eventually to reheat the universe. 
There will be Nambu-Goldstone bosons (NGB) present from the Inf sector due to spontaneous breaking
of the global symmetry as we discussed before. Their derivative type of coupling therefore
indicates that the inflaton system could decay into those goldstons with a decay width
$\Gamma_{{\rm {inf}} \rightarrow NSB} \simeq \frac{1}{64 \pi} \frac{m^3_{\chi}}{\Lambda^2}$. 
These particles would be produced
copiously during reheating and their abundance could spoil the success of  big-bang
nucleosynthesis (otherwise with massless NGB, it could just be  part of the radiation component
of the universe). Here we prescribe a resolution of this cosmological problem and show that 
effectively the reheating will take place via {\it {ISS mediation}}.

Without spoiling our description of the inflation sector, we perform a small deformation of
the set-up by choosing $\hat{m}_Q =$ diag$(m, \dots, m, m_Q)$ with $m \ll m_Q$ in $N_f = N +1$ case 
to we end up with an extra  $m \Lambda_0$Tr$T$ term in the superpotential of Eq.(\ref{wm}).
 In the limit $m \rightarrow 0$, it can be shown that the supersymmetric minimum coincides with
$B \bar B =0,$ det$\langle T \rangle = m_Q\Lambda^3_0, ~S =0$.
The insertion of the $m \Lambda_0 \chi$ term in the superpotential
of Infl sector will induce a tadpole term
for $S$ in the scalar potential that would shift the vev of $S$ from zero to
$\langle S \rangle \simeq m \langle \chi \rangle/m_Q$ at the end of inflation.
The inclusion of this new mass term for quarks would break the chiral symmetry
in the Infl sector explicitly and we would expect a mass term for the pseudo-NGBs. Following an analogy with the pion mass and using
a variant of Dashen formula\cite{dashen}, we can
argue that the these pseudo-NGBs will get a mass,
\be
m^2_t = O(1) m \frac{\Lambda^3_0}{\langle \chi \rangle^2}.
\ee
In order to forbid the decay of the inflaton into the NGBs kinematically we impose the constraint,
$m_{inf} < 2m_t$, namely 
\be
 O(1) \frac{m^2_Q}{\Lambda_0} < m < m_Q,
\ee
where we have included the fact that $m_Q < \Lambda_0$ as turned out from our analysis. We conclude that 
$10^{-4} < m/m_Q < 1$ (see Table 1) resolves the problem of NGBs. 

Then we see that the inflaton decays into the magnetic quarks $q, \tilde q$ of the dISS sector,
\be
V\ni \vert \frac{\partial W}{\partial \Phi}\vert^2 = \vert
q\tilde q + f_{2,5}\frac{\chi^4 \Lambda^4_0}{M^7_P} \Lambda_s\vert^2.
\ee
Since part of these $q, \tilde q$ are charged under $SU(5)$ of MSSM after gauging, the particular 
decay mode $\chi \rightarrow q\tilde q$
is instrumental for the production of MSSM particles through their
subsequent annihilation. 
The corresponding decay width is therefore given by
\be
\Gamma_{{\rm {inf}} \rightarrow q\tilde q} = {\frac{5}{8\pi}} \Bigl ({\frac{\mu^2}{\langle \chi \rangle}}\Bigr )^2
\frac{1}{m_{\rm {inf}}}.
\label{decay}
\ee
This particular way of reheating (we phrase it as {\it {ISS mediation}}) 
is a general feature of our scenario. This reheating mechanism is a new feature of our scenario\footnote{
We are aware that mediation is not mandatory 
for reheating, see, {\it e.g.}, \cite{majumdar}.}. The reheating temperature is given in Table 1.   
As one can see, the reheat temperature is very sensitive to the details of the models although it is always 
larger than the electroweak scale.

\section{Conclusions}
The origin of the supersymmetry breaking scale is mysterious as no precise model has been derived so far after more than 25 years of intensive
efforts. This lack of understanding of the supersymmetry breaking mechanism plagues the possibility of carrying out a predictive comparison of the sparticle spectrum
with present and future particle data coming from LEP and soon the LHC. Indeed, depending on the breaking scale many methods can be envisaged
in order to mediate the supersymmetry breaking to the observable sector. Two main  scenarios offer widely different results, gravity mediation requires a large breaking scale while gauge mediation can accommodate a much smaller scale. The latter allows a description at low energy without the need to invoke what happens close to the Planck scale (or the GUT scale). The absence of a well founded description plagues the analysis of inflation too. There again despite 25 years of efforts, hundreds of models have been proposed although none can be deemed as fully fundamental and problem-free. Recently, supergravity models derived from string theory following the original proposal by KKLT\cite{KKLT} and KKLLMT\cite{KKMT} have tried to tackle both inflation and supersymmetry breaking at the same time. Unfortunately, the simplest models require a large gravitino mass compared to the Hubble rate as well as a certain amount of fine-tuning\cite{bau}. It is nevertheless compelling that both inflation and supersymmetry breaking can be treated within the same framework.

Recently ISS\cite{ISS} have proposed that  supersymmetry breaking could occur in a long-lived metastable state. One salient point of their analysis is the fact that supersymmetry breaking is a low energy phenomenon occurring in the low energy regime of a SQCD theory. The transmission to the observable sector has been analysed in \cite{abel2} where gauge mediation appears as a natural candidate, only relying on low energy physics too. From the effective theory point of view, these facts are  very appealing as  both supersymmetry breaking and its mediation can be described at low energy.
In this paper we extend this analysis to the inflationary sector by requiring that it can also be described by the low energy effective theory of a SQCD theory. More explicitly, we have focused on an inflationary sector with 4 colours and 5 flavours. As usual,
the inflation scale is derived from the COBE bound on the CMB anisotropies. As advocated in \cite{ss}, the scale of supersymmetry breaking can be linked to the inflationary scale. This happens when the two sectors are coupled via gravitational interactions. As a result we find that the supersymmetry breaking scale can be low enough for gauge mediation. We also find  that reheating
at the end of inflation is obtained via  the decay of the inflaton in the quarks of the ISS sector. In conclusion, supersymmetry breaking is mediated from the inflationary sector and the reheating of the universe is due to the coupling of the inflaton with the magnetic quarks in the supersymmetry breaking ISS sector.

Of course, it is also relevant that both the supersymmetry breaking and inflationary sectors have an ultra violet completion. 
Moreover, this completion appears to be gauge theories with non-Abelian interactions. An interesting extension of our work 
would be to analyse the embedding of both the ISS sector and the inflation sector in string theory. This should be describable 
within the brane engineering framework\cite{kut2,kut1}. In such a description, both inflation as a result of brane motion and 
the interaction between the ISS and the inflation sectors should have a direct interpretation. This is left for future work. \\ \\

\noindent {\bf {Acknowledgements}} -  This work is supported by 
the RTN European Program MRTN-CT-2004-503369, the French ANR Programs PHYS@COL\&COS and 
DARKPHYS.  We thank the EU Marie Curie Research \& Training network `UniverseNet' 
(MRTN-CT-2006-035863) for support.


\end{document}